\documentclass[aps,prb,showpacs]{revtex4}
\usepackage{amssymb}
\usepackage{amsmath}
\usepackage{graphicx}
\usepackage{epsfig}

\begin{document}

\title{Cooling of a Micro-mechanical Resonator by the Back-action of Lorentz
Force}
\author{Ying-Dan Wang}
\author{K. Semba}
\author{H. Yamaguchi}
\affiliation{NTT Basic Research Laboratories, NTT Corporation, 3-1, Morinosato Wakamiya,
Atsugi-shi, Kanagawa 243-0198, Japan}
\date{\today }

\begin{abstract}
Using a semi-classical approach, we describe an on-chip cooling
protocol for a micro-mechanical resonator by employing a
superconducting flux qubit. A Lorentz force, generated by the
passive back-action of the resonator's displacement, can cool down
the thermal motion of the mechanical resonator by applying an
appropriate microwave drive to the qubit. We show that this on-chip
cooling protocol, with well-controlled cooling power and a tunable
response time of passive back-action, can be highly efficient. With
feasible experimental parameters, the effective mode temperature of
a resonator could be cooled down by several orders of magnitude.
\end{abstract}

\maketitle


\section{Introduction}
The rapid development of nano-technology has enabled fabrication of
micro or nano-mechanical
resonator~\cite{Clelandbook2002,Blencowe2004} with high frequency,
low dissipation and small mass. In the way to observe quantized
mechanical motion~\cite{LaHaye2004}, thermal fluctuation has become
a major obstacles. The limited cooling efficiency and poor heat
conduction at the milli-Kelvin temperatures of cryogenic
refrigerators has stimulated a number of studies on the active
cooling of micromechanical resonator (MR) in both classical regime
and quantum
regimes~\cite{Cohadon1999,Kleckner2006,Poggio2007,Metzger2004,Gigan2006,
Arcizet2006,Schliesser2006,Naik2006,Hopkins2003,Wilson2004,Martin2004,
Zhang2005,Brown2007,Xue2007b,Corbitt2007,Thompson2007,Lowry2008,You2008}.
Of these proposals, the optomechanical cooling is the most highly
developed one experimentally. For example, it has been successfully
used to demonstrate cooling~\cite{Thompson2007,Poggio2007} from room
temperature down to $6.82$ mK and from $2.2$ K down to $2.9$ mK.
Optomechanical cooling has also been studied theoretically in the
quantum limit~\cite{Marquardt2007,Genes2007,Wilson2007}.

In a thermal environment, a MR randomly vibrates around its
equilibrium position due to thermal fluctuation. The effective
temperature of a MR mode can be defined by the mean kinetic energy
of this mode. Cooling of MR is equivalent to suppressing the mean
amplitude of the Brownian motion, which is the major barrier to
precise displacement measurement. The optomechanical cooling of an
MR is achieved by the passive back-action or active feedback of such
optical forces as bolometric force~\cite{Braginsky2002,Metzger2004}
and radiation pressure
~\cite{Cohadon1999,Kleckner2006,Poggio2007,Gigan2006,Arcizet2006,Schliesser2006}
from a laser driven optical cavity.

Instead of employing an optical component, in this paper, we present
a cooling protocol for an MR in an on-chip superconducting circuit
containing three or four Josephson junctions (JJ). The
three-junction superconducting loop forms a flux
qubit~\cite{Mooij1999,Orlando1999,Wilhelm2006}. The typical energy
scale of the Larmor frequency of the flux qubit is normally several
GHz which is much larger than the oscillation frequency of the MR
studied in this paper. The persistent current in this loop exerts a
Lorentz force on an MR under an in-plane magnetic
field~\cite{Xue2007,Gaidarzhy2005}. This circulating superconducting
current is modulated by the thermal motion of the MR through
nonlinear Josephson inductance in a delayed way. Thus the Lorentz
force exerted on the MR depends on the motion of the MR itself. This
force acts as a passive back-action on the MR as the radiation
pressure in the optomechanical cooling strategy. The thermal motion
of the MR is damped by this back-action force with appropriate
parameters. To use a microwave bias to drive the superconducting
circuit helps to take away the thermal energy of the resonator. In
other words, in the linear response regime, a microwave drive
together with a flux qubit in a dissipative environment can be
treated as an effective bath whose temperature is lower than that of
the environmental temperature $T_0$. Thereby the mode temperature of
the resonator, which is proportional to the mean kinetic energy of
the fundamental oscillation mode, is
decreased~\cite{Martin2004,Clerk2005,Blencowe2005}. The strong and
tunable coupling between superconducting current and the MR
displacement is favorable as regards obtaining highly-efficient and
well-controlled cooling. Based on feasible experimental parameters,
we also provide a detailed analysis of the cooling efficiency of
this scheme. The estimation shows that it could constitute a
promising alternative to optomechanical cooling.

This paper is organized as follows: In Sec. II, we describe the
setup we use to implement our cooling scheme. Sec. III presents the
intuitive picture of this back-action cooling by Lorentz force,
together with a general formalism for dealing with self back-action
cooling. In Sec. IV, the cooling efficiency of this physical system
is determined based on a detailed calculation of the "spring
constant" and the response time of the Lorentz force. The lengthy
calculation part concerning to the master equation and its
steady-state solution are given in the Appendix. In Sec. V, cooling
efficiency is estimated by using feasible experimental parameters .
Finally in Sec. VI, we discuss the possible advantages, fluctuation
and measurement protocol of this cooling protocol.

\section{The Setup}
\begin{figure}[tp]
\begin{center}
\includegraphics[bb=106 450 490 570,scale=0.6,clip]{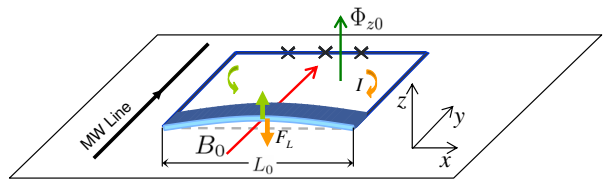}
\end{center}
\caption{(Color on line) Schematic diagram of our setup. A
doubly-clamped mechanical beam is incorporated in a superconducting
loop with a superconducting Josephson junction (indicated with
crosses) flux qubit. The initial bias in the loop is controlled by
magnetic flux $\Phi _{z0}$ in $z$ direction. A coupling magnetic
field $B_{0}$ is imposed on the beam in the $y$ direction and the
supercurrent under this magnetic field imposes a Lorentz force on
the beam. A microwave line introduces a microwave bias to the qubit
loop. The magnitude of the Lorentz force depends on the motion of
the beam through the change of the supercurrent in the loop in a
delayed way. This delayed back-action damps the motion of the beam
in the $z$ direction and thereby cools down the thermal motion of
the beam.} \label{fig:circuit}
\end{figure}

A schematic diagram of the setup for this cooling protocol is shown
in Fig.\ref{fig:circuit}.
In the $x$-$y$ plane, a doubly-clamped micro-mechanical beam with an
effective length $L_{0}$ is incorporated in a superconducting loop
with three small-capacitance Josephson junctions. This mechanical
beam can be created from a surface-micromachined silicon or carbon
nanotube~\cite{Poncharal1999} coated with superconducting material,
or a self-supporting metallic airbridge. The fundamental vibration
mode of the beam can be well approximated by using a harmonic
resonator with oscillation frequency $\omega _{b}$. With a proper
bias magnetic flux, two classical stable states of the 3-JJ loop
carry persistent currents in opposite directions. There is a finite
tunneling rate $\Delta$ between the two classical persistent-current
states (throughout this article, we let $\hbar =1$). By choosing
parameters carefully, the subspace spanned by the two states is well
separated from other energy levels and the superconducting loop with
Josephson junctions forms a two-level system (superconducting flux
qubit)~\cite{Mooij1999,Orlando1999} and coherent dynamics can be
observed~\cite{Chiorescu2003,Saito2004}. Recently, the
microwave-induced cooling of a flux qubit has been demonstrated
experimentally~\cite{Valenzuela2006} assisted by the third energy
eigen-state of the three-junction superconducting loop. The qubit
ground state $|g\rangle $ and excited state $|e\rangle$ are coherent
superpositions of two persistent current states denoted by
$|0\rangle$ (clockwise current state) and $|1\rangle$
(anti-clockwise current state). The energy spacing between the two
eigenstates is $\Omega =\sqrt{\Delta ^{2}+\varepsilon ^{2}}$ where
$\varepsilon =2I_{p}\left( \Phi _{ext}-\Phi _{0}/2\right) $ is the
energy spacing of the two classical current states ($|0\rangle$ and
$|1\rangle$), with $\Phi _{ext}$ the external magnetic flux through
the loop, $I_{p}$ the largest persistent current in the loop and
$\Phi _{0}$ the flux quantum. A microwave line is placed close to
the circuit and generates microwave drive with frequency $\omega
_{d}$ on the flux qubit. Under the coupling magnetic field $B_{0}$
along $y$ direction, the persistent superconducting current
generates a Lorentz force $F_{L}$ on the MR along $z$ direction.
This force couples the flux qubit with the oscillation motion of the
MR. In the quantum regime of the MR, this configuration provides a
solid-state analog of cavity QED system in strong coupling
limit~\cite{Xue2007}. Similar setting with a coupled MR and dc-SQUID
was proposed recently to study the displacement detection and
decoherence of mechanical
motion~\cite{Zhou2006,Buks2006,Blencowe2007}

In the present paper, we concentrate on a different regime where the
oscillation frequency of the MR is much smaller than the Larmor
frequency of the flux qubit so that the MR can be treated as a
classical harmonic oscillator. Since the qubit and the MR energy
scales differs greatly, the dynamics of the composite system can be
handled following the line of the Born-Oppenheimer approximation:
The master equation of the qubit is established by assuming a
certain MR displacement $z$; the steady-state solution for the qubit
dynamics is inserted into the classical Langevin equation of the MR
to derive the noise spectrum of the mechanical displacement.

\section{The cooling mechanism}

In this section, we first present an intuitive understanding of the
cooling protocol based on the Lorentz force back-action. Using the
classical Langevin equation of the MR, we then proceed to analyze
how this back-action leads to the suppression of the random motion
of the MR. The explicit form of the Lorentz force back-action in the
Langevin equation will be derived in the next section.

The composite flux qubit and MR system is kept at an environmental
temperature $T_{0}$ in a dilution refrigerator. The initial constant
bias flux of the loop is $\Phi_{z0}$ in the $z$ direction. When the
MR is displaced to $z(t)$ by thermal noise, the total magnetic flux
in the superconducting loop is changed to $\Phi
_{z0}+B_{0}L_{0}z(t)$. In principle, for an MR with an aspect ratio
close to 1, thermal fluctuation also induces oscillation in the $y$
direction. However, since the magnetic field in the $z$ direction is
much smaller than the coupling magnetic field $B_0$ in the $y$
direction, in this work we only study the motion in the $z$
direction. An increase or decrease in magnetic flux leads to a
change in the persistent current $I$ in the loop after the system
reaches a metastable state after a response time delay
$\tau_\text{resp}$. Owing to the nonlinearity of Josephson junction
inductance, the supercurrent in a metastable state depends on the
total bias magnetic flux and hence depends on the displacement of
the MR. Since the Lorentz force is proportional to the supercurrent,
the Lorentz force exerted on the MR $F_{L}(t+\tau_\text{resp})$
depends on the displacement of the MR $z(t)$ before the time delay.
This means that there exists a passive back-action mechanism for the
MR: the motion of the MR leads to a delayed force on the MR itself.
We found that, with a proper microwave drive (red-detuned with the
qubit), this passive back-action damps the thermal motion of the MR.
This cooling protocol is similar to that of the self cooling
experiments based on optomechanical
coupling~\cite{Braginsky2002,Metzger2004}.

The above intuitive picture can be clearly understood by studying
the MR dynamics. To accomplish this, we start by establishing an
equation of motion for the MR.

The coupling term in the Hamiltonian of qubit-resonator composite
system is~\cite{Xue2007}
\begin{equation}
H_{\text{int}}=B_{0}L_{0}\hat{I}z.
\end{equation}
where $\hat{I}$ is the current operator. Then the Lorentz force on
the resonator is
\begin{equation}
\hat{F}_{L}=-\frac{\partial H_{\text{int}}}{\partial
z}=-B_{0}L_{0}\hat{I}
\end{equation}
Suppose the MR is displaced at $z$, after a time delay
$\tau_\text{resp} $, the resulting Lorentz force in the z metastable
state $F_{L}^{\left( s\right) }\left( z\right) \equiv
\mathrm{Tr}\left( \rho _{q}^{\left( s\right) }\left( z\right)
\hat{F}_{L}\right) $ depends on $z$, where $\rho _{q}^{\left(
s\right) }$ is the reduced density matrix of the qubit in a
metastable state. For a small displacement $z$, the Lorentz force
can be expanded to the linear order of $z$ as $F_{L}^{\left(
s\right) }\left( z\right) =F_{0}+k_{L}z$ with $k_{L}$ the effective
"spring constant" of the Lorentz force. Since this force is a
delayed response to the motion of the MR, its effect on the MR can
be described by a delayed response function $h\left(t-t^{\prime
}\right) \equiv 1-e^{-\gamma (t-t^{\prime })}$ with
$\gamma=1/\tau_\text{resp}$. Under the Lorentz force, the equation
of motion for the MR mode with mass $m$, rigidity
$k=m\omega_{b}^{2}$ and an inherent damping rate $\Gamma$ is as
follows~\cite{Metzger2004}
\begin{equation}
m\frac{d^{2}z}{dt^{2}}+m\Gamma
\frac{dz}{dt}+kz=F_{th}+\int_{0}^{t}\frac{dF_{L}^{\left( s\right)
}\left[ z\left( t^{\prime }\right) \right] }{ dt^{\prime }}h\left(
t-t^{\prime }\right) dt^{\prime },  \label{mq}
\end{equation}
where $F_{th}$ is the Brownian fluctuation force which is related to
the environmental temperature $\left\langle F_{th}\left( t\right)
F_{th}\left( t^{\prime }\right) \right\rangle =2k_{B}T_{0}m\Gamma
\delta \left( t-t^{\prime }\right) $. The stable solution of the
motion equation (\ref{mq}) is related to the effective mode
temperature $T_{\text{eff}}$ of the MR by the equipartition theorem
$k_{\text{eff}}\left\langle z^{2}\right\rangle =k_{B}T_{\text{eff}}$
where $k_{B}$ is the Boltzman constant. Then the cooling efficiency
$\eta =T_{0}/T_{\text{eff}}$ can be written as
\begin{equation}
\eta^{-1}=\frac{\Gamma \omega_{\text{eff}}^{2}}{\pi}\int_{-\infty
}^{+\infty }\frac{d\omega}{\left( \omega _{\text{eff}}^{2}-\omega
^{2}\right) ^{2}+\Gamma _{\text{eff}}^{2}\omega ^{2}},  \label{effi}
\end{equation}
where
\begin{eqnarray}
\Gamma _{\text{eff}} &=&\Gamma +\Gamma _{1},  \notag \\
\omega _{\text{eff}}^{2} &=&\omega _{b}^{2}\left( 1-\frac{\gamma
^{2}}{\omega ^{2}+\gamma ^{2}}\frac{k_{L}}{k}\right) .
\end{eqnarray}
and
\begin{equation}
\Gamma _{1}=\Gamma Q_{M}\frac{\omega _{b}\gamma }{\omega ^{2}+\gamma
^{2}}\frac{k_{L}}{k},
\end{equation}
with $Q_{M}=\omega _{b}/\Gamma $ is the quality factor of the MR.
For $\left\vert k_{L}\right\vert \ll k$, the integral can be
explicitly carried out as~\cite{Metzger2004}
\begin{equation}
\eta=1+Q_{M}\frac{k_{L}}{k}\frac{\omega _{b}\gamma }{\omega _{b}^{2}+\gamma
^{2}}.  \label{eff2}
\end{equation}

From Eq.(\ref{eff2}), it can be seen that the sign of $k_{L}$
determines whether  the resonator is cooled or heated, i.e. when
$k_{L}$ is positive, the Lorentz force damps the motion of the MR
and the final temperature is lower than the original one, and vice
versa. For a positive $k_{L}$, the cooling efficiency increases
linearly with $k_{L}$. However, it should be noticed that both the
linear response regime and the definition of a single-mode resonator
are valid only for $k_L/k\ll 1$.

\begin{figure}[bp]
\begin{center}
\includegraphics[bb=127 293 434 532,scale=0.7,clip]{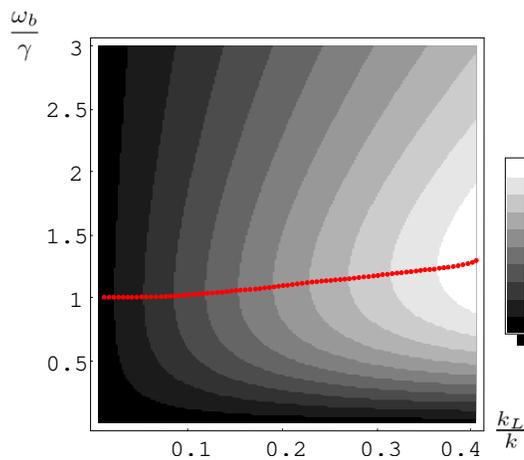}
\end{center}
\caption{(Color on line) The dependence of the cooling efficiency on
the ratios of $k_{L}/k$ and $\protect\omega _{b}/\protect\gamma $.
The magnitude of the cooling efficiency is re-scaled and indicated
by the gray level (higher efficiencies are shown lighter). The other
parameters are the same as those used for estimating the cooling
efficiency in Sec. V. The red dots indicates the point with the
largest efficiency at each given $k_{L}/k$ value.}
\label{fig:effi_k_ome}
\end{figure}
The cooling efficiency is shown by a contour plot in Fig.
\ref{fig:effi_k_ome} as a function of two ratios $\omega _{b}/\gamma
$ and $k_{L}/k$. $\omega _{b}/\gamma $ characterizes the ratio
between the time scale of the response time and the oscillation
period while $k_{L}/k$ is the scaled cooling strength of the passive
Lorentz force. The magnitude of the efficiency is indicated by the
gray level: Higher efficiency is represented with a lighter color.
It is seen that the cooling efficiency increases with $k_{L}/k$. The
strong coupling in the solid state cavity QED makes it promising to
achieve larger $k_{L}$. For a given $k_{L}/k$, the largest cooling
efficiency can be achieved by optimizing $\omega _{b}/\gamma$. The
optimal points for each $k_L/k$ value are indicated by the red dots
in Fig. \ref{fig:effi_k_ome}. The optimal $\omega _{b}/\gamma $
slightly increases with $k_{L}$. For $k_{L}/k\ll 1$, which is in the
case of interest, the red dots indicate that the optimal cooling is
realized for $\omega _{b}=\gamma$. This means the largest cooling
efficiency is achieved when the back-action response time matches
the oscillation period of the resonator. As we show later, the
response time is of the order of the relaxation time of the flux
qubit. The qubit relaxation rate ranges from sub MHz to several tens
of MHz, depending on the operating point $\varepsilon
_{0}$~\cite{Yoshihara2006,Kakuyanagi2007}, this implies that the
flux qubit loop is suitable for cooling a resonator over broad
frequency range. Moreover, since the operating point $\varepsilon
_{0}$ can be controlled by magnetic flux $\Phi _{z0}$, the
back-action response time is tunable \textit{in situ}.

\section{Calculation of the back-action of the Lorentz force}

As we discussed in the previous section, without considering
additional fluctuation, the damping effect of the Lorentz force
leads to cooling or heating depending on the sign of the effective
spring constant $k_{L}$: A positive $k_{L}$ leads to cooling while a
negative $k_{L}$ results in heating. The cooling efficiency is
proportional to the magnitude of $k_{L}$. Before proceeding with our
discussion of the cooling efficiency in realistic experiment, it is
necessary to obtain an explicit expression of $k_L$ from the
dynamics of the flux qubit system.

The Lorentz force in a metastable state for a given displacement $z$
is
\begin{equation}
F_{L}^{\left( s\right) }\left( z\right)
=-B_{0}I_{p}L_{0}\left\langle \sigma _{z}\right\rangle _{s}
\end{equation}
where $\left\langle \sigma _{z}\right\rangle _{s}$ is the
expectation value of $\sigma _{z}$ in a steady-state. And $\sigma_z$
and $\sigma_x$ are Pauli matrixes defined by the persistent current
states as:
\begin{eqnarray}
\sigma_z=|0 \rangle\langle 0|-|1 \rangle\langle 1| \notag \\
\sigma_x=|0\rangle\langle 1|+|1 \rangle\langle 0|
\end{eqnarray}
$\left\langle \sigma _{z}\right\rangle _{s}$ can be computed from
the dynamics of the driven flux qubit in a dissipative environment.
To do this, we start with the effective Hamiltonian of the qubit at
a given displacement $z$ of the MR
\begin{equation}
H_q=\frac{\varepsilon \left( z\right) }{2}\sigma _{z}+\frac{\Delta
}{2}\sigma _{x}+A\sigma _{z}\cos \omega _{d}t.  \label{H1}
\end{equation}
Here $A$ characterizes the amplitude of the microwave drive,
\begin{equation}
\varepsilon \left( z\right) =\varepsilon _{0}+gz
\end{equation}
with $g=2B_{0}I_{p}L_{0}$, $\varepsilon _{0}=2I_{p}\Phi _{z0}$
denotes the initial bias away from the degeneracy point. If the
qubit is biased at the degeneracy point, $\varepsilon _{0}=0$.

By defining a new set of Pauli operators
\begin{eqnarray}
\sigma _{z}^{\prime } &=&\sigma _{z}\cos \theta \left( z\right) +\sigma
_{x}\sin \theta \left( z\right),  \notag \\
\sigma _{x}^{\prime } &=&-\sigma _{z}\sin \theta \left( z\right) +\sigma
_{x}\cos \theta \left( z\right),  \label{t1}
\end{eqnarray}
we diagonalize the first two terms of eq.(\ref{H1}) as,
\begin{equation}
H_q=\frac{\Omega \left( z\right) }{2}\sigma _{z}^{\prime }+A\left( \sigma
_{z}^{\prime }\cos \theta \left( z\right) -\sigma _{x}^{\prime }\sin \theta
\left( z\right) \right) \cos \omega _{d}t,  \label{h2}
\end{equation}
where
\begin{equation}
\cos \theta \left( z\right) =\frac{\varepsilon \left( z\right)
}{\Omega (z)}\ ,\ \ \sin \theta \left( z\right) =\frac{\Delta
}{\Omega \left( z\right) }\ ,
\end{equation}
and
\begin{equation}
\Omega \left( z\right) =\sqrt{\Delta ^{2}+\varepsilon ^{2}\left(
z\right)}\approx \Omega _{0}+\tilde{g}z,
\end{equation}
with $\tilde{g}=g\varepsilon _{0}/\Omega _{0}$, $\Omega
_{0}=\sqrt{\Delta ^{2}+\varepsilon _{0}^{2}}$ is the energy spacing
between the qubit eigenstates when the displacement $z$ of the MR is
zero. Here we have used the fact that the displacement $z$ is very
small and $\cos\omega _{d}t$ is a fast-oscillating term.

If the drive is near-resonant with the qubit frequency $\Omega
_{0}$, by performing unitary transformation $U_{R}=\exp (i\sigma
_{z}^{\prime }\omega _{d}t/2)$, the Hamiltonian (\ref{h2}) in the
rotating frame is transformed to
\begin{equation}
H_{R}=i\left( \frac{d}{dt}U_{R}( t)\right) U_{R}^{\dag
}(t)+U_{R}\left( t\right) HU_{R}^{\dag }(t)=\frac{\delta \left(
z\right) }{2}\sigma _{z}^{\prime }+\frac{A^{\prime}(z)}{2}\sigma
_{x}^{\prime}  \label{t2}
\end{equation}
with $A^{\prime }\left( z\right) =-A\sin \theta \left( z\right) $,
$\delta \left( z\right) =\delta \omega +\tilde{g}z$ and $\delta
\omega =\Omega _{0}-\omega _{d}$ is the detuning between the qubit
free energy and the drive. The term $A\sigma _{z}^{\prime }\cos
\theta \left( z\right) \cos \omega _{d}t$ and
$A\sin\theta(z)\sigma_+ e^{i\omega_dt}+h.c$ in eq. (\ref{h2}) is
neglected because $\exp(\pm i\omega _{d}t)$ and $\exp(\pm 2i\omega
_{d}t)$ are fast-oscillating terms since the drive frequency is
close to the qubit energy spacing. The diagonalized $H_{R}$
\begin{equation}
H_{R}=\frac{\omega _{0}\left( z\right) }{2}\tilde{\sigma}_{z}
\end{equation}
is expressed by another set of Pauli operators
\begin{eqnarray}
\tilde{\sigma}_{z} &=&\sigma _{z}^{\prime }\cos \beta \left( z\right)
+\sigma _{x}^{\prime }\sin \beta \left( z\right),  \notag \\
\tilde{\sigma}_{x} &=&-\sigma _{z}^{\prime }\sin \beta \left( z\right)
+\sigma _{x}^{\prime }\cos \beta \left( z\right),  \label{t3}
\end{eqnarray}
with
\begin{equation}
\sin \beta \left( z\right) =\frac{A^{\prime }\left( z\right) }{\omega
_{0}\left( z\right) },\ \ \cos \beta \left( z\right) =\frac{\delta \left(
z\right) }{\omega _{0}\left( z\right) },
\end{equation}
and
\begin{equation}
\omega _{0}\left( z\right) =\sqrt{\delta ^{2}\left( z\right) +A^{\prime
2}\left( z\right)}\ .  \label{omega0}
\end{equation}

With the above transformation relations, we can derive the master
equation for this dissipative flux qubit in a rotating frame.
Solving the master equation to obtain its steady-state solution and
using the above relations to transform it back to an experimental
frame, we can obtain $\langle\sigma_z\rangle_s$ in an experimental
frame (see Appendix for detail)
\begin{equation}
\left\langle \sigma _{z}\right\rangle _{s}=\frac{2\cos ^{2}\beta
\left( z\right) }{1+\cos ^{2}\beta \left( z\right) }\cos \theta
\left( z\right) \label{sz}
\end{equation}
From the last equation of (\ref{ds}), the response time needed for
the supercurrent to reach this value is
\begin{equation}
\tau_\text{resp}=\frac{1}{\gamma}
=\frac{1}{\Gamma_0}\frac{2}{1+\cos^2\beta_0}  \label{gamma1}
\end{equation}
with $\cos\beta_0\equiv\cos\beta(z=0)$. The response time is of the
order of the qubit relaxation time. Shifting the qubit bias will
change the qubit relaxation time and hence change the response time.
On the other hand, as found with SET-assisted cooling
\cite{Clerk2005}, the response time also depends on the drive
detuning and power. This is in contrast to optomechanical cooling
where the response time is always determined by the cavity ring-down
time.

Since $z$ is small, the force can be expanded to the linear order of
$z$: $F_{L}^{\left( s\right) }\left( z\right) =F_{0}+k_{L}z$, where
\begin{equation}
F_{0}=-B_{0}I_{p}L_{0}\frac{2\cos ^{2}\beta _{0}}{1+\cos ^{2}\beta _{0}}\cos
\theta _{0},
\end{equation}
and the effective spring constant of the Lorentz force is
\begin{equation}
k_{L}=B_{0}I_{p}L_{0}\left( \frac{\partial \left\langle \sigma
_{z}\right\rangle _{s}}{\partial z}\right) _{z=0}.
\end{equation}
The derivation of the above equation can be explicitly carried out
as
\begin{equation}
k_{L}=\frac{\left( 2B_{0}I_{p}L_{0}\right) ^{2}}{\omega _{0}}\left(
\frac{\varepsilon _{0}^{2}}{\Omega _{0}^{2}}\frac{2\omega _{0}\delta
\omega A^{2}}{\left( \omega _{0}^{2}+\delta \omega ^{2}\right)
^{2}}+\frac{\omega _{0}}{\Omega _{0}}\frac{\Delta ^{2}}{\Omega
_{0}^{2}}\frac{\delta \omega ^{2}}{\omega _{0}^{2}+\delta \omega
^{2}}\right)   \label{kl}
\end{equation}
where $\omega _{0}$ is the value of eq.(\ref{omega0}) at $z=0$:
\begin{equation}
\omega _{0}=\sqrt{\delta \omega ^{2}+A^{2}\frac{\Delta ^{2}}{\Omega
_{0}^{2}} }.
\end{equation}
When the flux qubit is biased near the degeneracy point,
$\varepsilon _{0}\lesssim \Delta $, in the bracket of eq.
(\ref{kl}), the second term is much smaller than the first one owing
to the small prefactor $\omega _{0}/\Omega _{0}$ (in this case,
$\Omega _{0}$ is about several gigahertz while the detuning and
drive amplitude are about several megahertz) and we neglect the
second term
\begin{equation}
k_{L}\approx \left( 2B_{0}I_{p}L_{0}\right) ^{2}\frac{\varepsilon
_{0}^{2}}{\Omega _{0}^{2}}\frac{2\delta \omega
A^{2}}{\left(\omega_0^{2}+\delta \omega ^{2}\right) ^{2}}.
\end{equation}

\begin{figure}[tp]
\begin{center}
\includegraphics[bb=137 319 509 537,scale=0.7,clip]{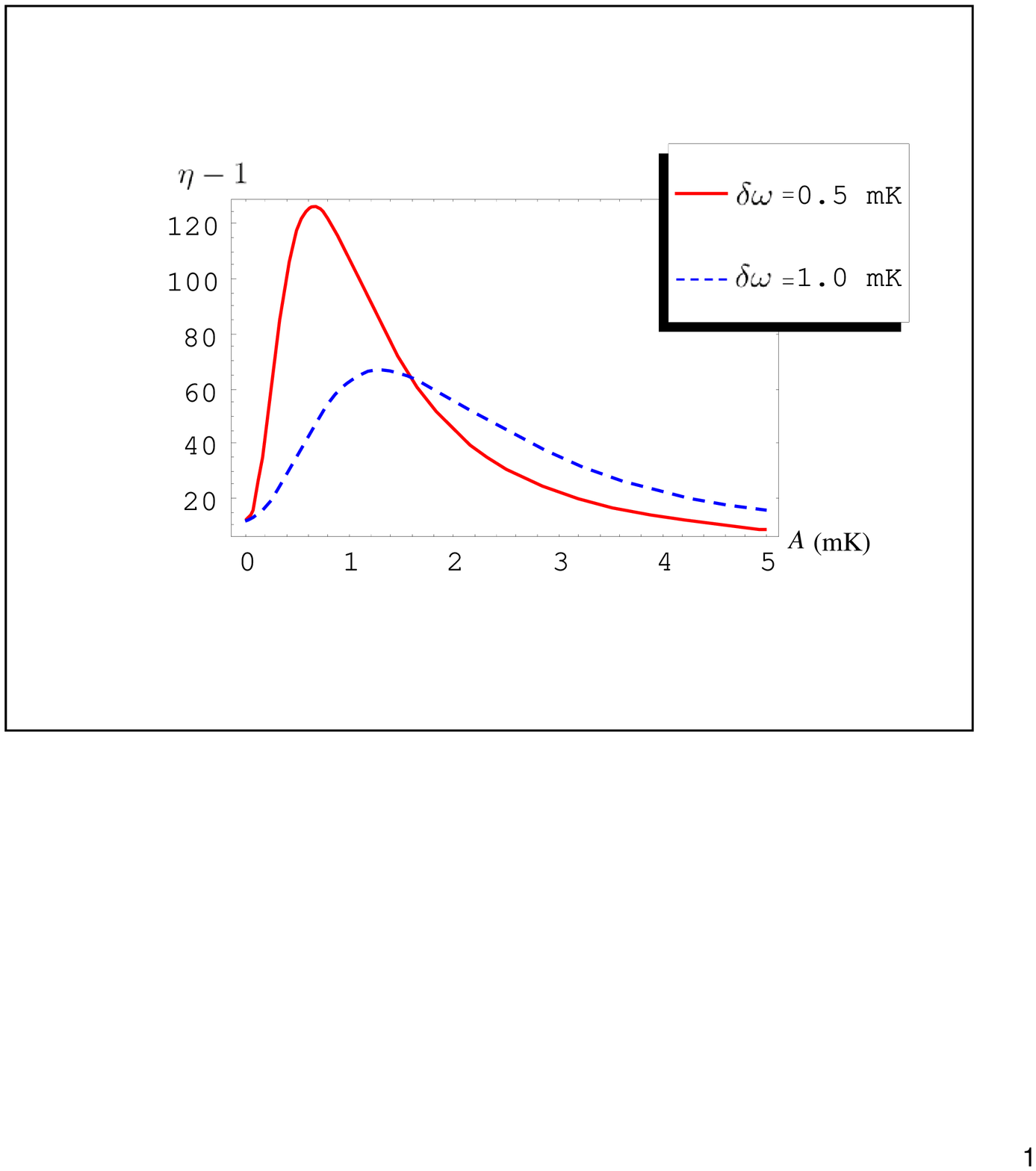}
\end{center}
\caption{(Color online) The dependence of the change in the cooling
efficiency $\eta-1$ on the drive amplitude $A$ with drive detuning
$\delta\omega=1$ mK and $\delta\omega=5$ mK respectively.
$\delta\omega$ and $A$ are in milli-Kelvin units. All the other
parameters are the same as those used to estimate of cooling
efficiency in Sec. V.} \label{fig:k_A}
\end{figure}

Eq. (\ref{kl}) represents the explicit form of $k_{L}$ expressed by
the physical quantities of this system. As we have already noticed,
the cooling efficiency is closely related to this effective spring
constant $k_{L}$. Using this expression, we are able to study the
cooling efficiency of this physical system. Inserting eqs.
(\ref{kl}) and (\ref{gamma1}) into eq. (\ref{eff2}), we obtain the
cooling efficiency of this system
\begin{equation}
\eta=1+\frac{\left( 2B_{0}I_{p}L_{0}\right) ^{2}}{m\omega _{b}\omega
_{0}}\frac{4Q_{M}\Gamma _{0}\left( 1+\cos ^{2}\beta _{0}\right)
}{4\omega _{b}^{2}+\Gamma _{0}^{2}\left( 1+\cos ^{2}\beta
_{0}\right) ^{2}}\left( \frac{\varepsilon _{0}^{2}}{\Omega
_{0}^{2}}\frac{2\omega _{0}\delta \omega A^{2}}{\left( \omega
_{0}^{2}+\delta \omega ^{2}\right) ^{2}}+\frac{\omega _{0}}{\Omega
_{0}}\frac{\Delta ^{2}}{\Omega _{0}^{2}}\frac{\delta \omega
^{2}}{\omega _{0}^{2}+\delta \omega ^{2}}\right) .  \label{eff3}
\end{equation}
$\eta-1$, i.e., the second term of the right part of the above
equation, is the change in the cooling efficiency caused by the
Lorentz force back-action. If $\eta-1$ is positive, the back-action
leads to cooling, and vice versa.
\begin{figure}[bp]
\begin{center}
\includegraphics[bb=119 290 524 531,scale=0.7,clip]{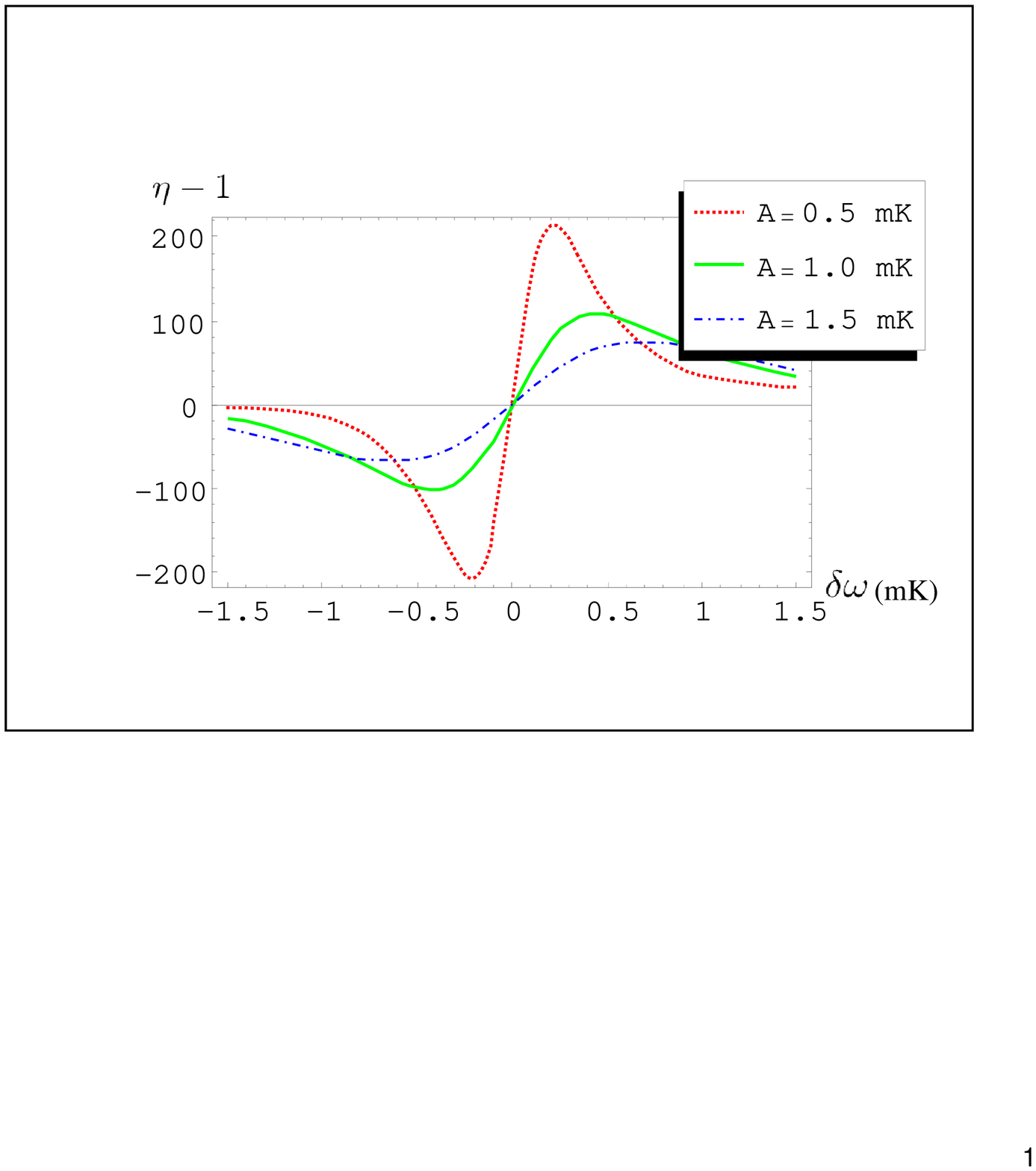}
\end{center}
\caption{(Color online) The dependence of the change in cooling
efficiency $\eta-1$ on the drive frequency $\delta\omega$ for
different drive amplitude. $\delta\omega$ and $A$ are in
milli-Kelvin units. All the other parameters are the same as those
used to estimate the cooling efficiency in Sec. V.} \label{fig:k_dw}
\end{figure}
The dependence of $\eta-1$ with respect to the drive detuning
$\delta\omega$ and amplitude $A$ near degeneracy point is shown in
Fig. \ref{fig:k_dw}. It can be seen that for a positive $\delta
\omega $ (red detuning), the cooling regime is reached, and vice
versa (however, note that $\eta<0$ corresponds to an unstable case
rather than heating). There is no cooling or heating for zero drive
$A=0$. This is consistent with optomechanical cooling. However,
because of the difference between SU(2) and Heisenberg algebra, for
$\delta \omega
>0$, the cooling efficiency does not scale with $A$ monotonically
but reaches its maximum at $A=\sqrt{2}\delta \omega$ as shown in
Fig. \ref{fig:k_A}. Physically, this is because too strong a drive
destroys the back-action mechanism owing to its dominant rapid
dynamics and even drive the qubit out of the two-level subspace. It
is also shown in Fig. \ref{fig:k_dw} that, as $A$ increases, the
variation in the cooling efficiency with respect to $\delta\omega $
around the peak tends to become much flatter. This feature makes the
cooling efficiency robust as regards operating point fluctuation by
increasing the drive amplitude. However, the largest cooling
efficiency also decreases with the increases in $A$. Therefore, in
real experiment, a trade-off is needed to obtain both stable control
and high cooling efficiency.

It is worth to mentioning that, if the qubit is biased at the
degeneracy point, the second term in eq. (\ref{kl}) cannot be
omitted. Thus we obtain a small but finite cooling effect even at
the degeneracy point. Based on the first-order perturbation theory,
side-band cooling is not possible at the qubit degeneracy point.
However, driven Rabi oscillation at degeneracy point has been
demonstrated experimentally~\cite{Ilichev2003}. This is explained by
fluctuation of the operating point or higher-order perturbation
theory~\cite{Hauss2007}. It is interesting to observe that this
effect is preserved by our semi-classical treatment of back-action
cooling. However, based on the treatment of quantized MR, other
resonant conditions may be needed if we are to study this
phenomenon~\cite{Hauss2007}.

\section{Cooling efficiency in practical systems}

Having obtained the effective spring constant of the Lorentz force
and established the relationship between the cooling effect and the
back-action mechanism, we can now estimate the cooling efficiency of
this protocol for a real system. To achieve higher cooling
efficiency, it is necessary to increase the mechanical quality
factor $Q_{M}$ and the coupling strength $k_{L}$. A higher quality
factor implies that the MR is less affected by the thermal
environment. Greater coupling strength means stronger back-action
damping. Moreover, as shown in Fig. \ref{fig:effi_k_ome}, another
crucial factor with respect to improving the cooling efficiency is
the ratio between the MR frequency and the flux qubit relaxation
rate, i.e., the time scales of the oscillation period and the
feedback response from the qubit. Efficient cooling is achieved when
the two time scales match. In self-cooling experiments with
optomechanical coupling, this is one of the main experimental
challenges~\cite{Karrai2006}. In our proposal, the response time
depends on the relaxation rate of the flux qubit. This rate can be
controlled from $0.1$ MHz to $10$ MHz by controlling the initial
bias magnetic flux $\Phi _{z0}$ in the $z$ direction. By controlling
the high frequency ($~\Delta/\hbar$) noise, e.g., by deliberately
attaching an external impedance to the circuit, we can change the
qubit spontaneous emission time. Hence the relaxation rate of the
flux qubit can be further monitored to match the frequency of the
mechanical mode. This feature enables us to optimize the cooling of
MR with different mode frequencies and greatly improves the cooling
power.

We made our estimation based on a flux qubit loop with an aluminum
Josephson junction. In this case, the relaxation time around the
degeneracy point of the qubit is of the order of one hundred
nanoseconds to several
microseconds~\cite{Yoshihara2006,Kakuyanagi2007}. Hence it is
suitable for a cooling MR mode with MHz frequency. For example, at
an environmental temperature of $300$ mK, a flux qubit with
$I_{p}=600$ nA, $\Delta =5$ GHz, $\varepsilon_0 =1$ GHz and the
relaxation rate close to the degeneracy point is $\gamma=5$ MHz. The
qubit is driven by a microwave with $\omega _{d}=5.089$ GHz and
$A=14.1$ MHz. When the driven qubit is used to cool a doubly-clamped
Si beams~\cite{Roukes2000} with $\omega _{b}=5$ MHz, $k=0.1$ N/m,
$L_{0}=5$ $\mu $m and quality factor $Q_{M}=10^{4}$, assuming
$B_{0}=5$ mT, we obtain $k_{L}=2.72\times 10^{-3}$ N/m. The cooling
efficiency $\eta $ is about $1.27\times 10^{2}$, which means the
effective mode temperature of the MR can be cooled to about $2.36$
mK. Since $k_{L}$ is proportional to the square of the coupling
magnetic field, increasing $B_{0}$ can greatly enhance the cooling
power. In principle, the upper limit of $B_{0}$ is the critical
magnetic field $B_{c}$ of the superconducting material in order to
preserve superconductivity. For aluminum, $B_{c}\approx 9.9$ mT. But
the in-plane magnetic field can be much stronger than this critical
value (e.g. even larger than $100$ mT~\cite{Ferguson2006}) without
destroying the superconductivity. In addition, suitable arrangement
of the magnetic field can result in the strong magnetic effect
acting only in the MR region while being largely canceled out at the
junction location. This means this cooling protocol could be
potentially more powerful than the above estimation.

However, one major problem with too strong an in-plane magnetic
field in the experiment is the possible tiny vibration of the sample
with respect to the coupling magnetic field. With a strong magnetic
field, a significant shift may be induced in the operating point by
the tiny vibration. One way to solve this problem is to design the
gradiometer-type flux qubit or to use an on-chip magnetic field
generating coil~\cite{Xue2007}.

\section{Discussions}

The basic idea behind our cooling protocol is similar to the
self-cooling of a micro-mirror by optomechanical
coupling~\cite{Karrai2006}: The flux qubit plays the role of the
optical cavity and the microwave drive on the qubit acts as the
laser drive on the cavity. The Lorentz force produces a passive
back-action on the resonator in the same way as photothermal force
or radiation pressure.

However, the mathematical treatment of a two-level artificial atom
is rather different from that of a bosonic field. Moreover, in
practice the two systems have distinct physical nature: (1) An
on-chip solid-state system without any optical component might have
certain advantages as regards its application; (2) The coupling
between a flux qubit and a resonator can be controlled by
controlling the applied magnetic field $B_{0}$ which is independent
of the qubit free Hamiltonian and microwave drive. This is different
from the coupling system of a Cooper pair box and NAMR
\cite{Irish2003} where the bias voltage $V_g$ modifies the coupling
coefficient as well as the free Hamiltonian of the charge qubit.
This feature gives the system more flexibility in terms of
increasing the cooling power and switching the cooling process on
and off ; (3) In order to near-resonant to laser, the frequency of
the FP cavity is more than $1$ THz. To achieve efficient
optomechanical cooling for a MHz oscillator, the optical quality
factor of the FP cavity should be about $10^{8}$. This calls for a
mirror with extremely high finesse. But in our case, it is easier to
match the two energy scales since the relaxation rate at the
degenerate point is typically of the order of microsecond. (4) The
relaxation rate of a flux qubit, which determines the back-action
response time, can be modified \textit{in situ} by the bias magnetic
flux in the $z$ direction. In principle with a GHz oscillator, we
can work in both the $\gamma \approx \omega _{b}$ and $\gamma \ll
\omega _{b}$ regimes for this Lorentz-force cooling. The latter
regime is desirable for cooling towards the quantum
limit~\cite{Marquardt2007}. This implies that our present proposal
might be able to cool the MR into its quantum ground state. However,
this question requires further discussion involving a consideration
of quantum fluctuation.

In the above discussion, we only considered the damping of the
resonator caused by the qubit via the Lorentz force. However,
according to the fluctuation-dissipation theorem, fluctuation is
always associated with dissipation. Additional fluctuation is also
induced by the back-action of the qubit system. For a more
comprehensive description, we need to take this fluctuation into
account. The additional fluctuation can be represented by an
effective temperature $T_1$. Therefore, the resonator is effectively
in contact with two different
baths~\cite{Martin2004,Clerk2005,Blencowe2005}, one is the real
environment with temperature $T_0$, and damping rate $\Gamma$ and
the other is the effective bath with effective temperature $T_{1}$,
and damping rate $\Gamma _{1}$. Physically, the effective bath is
formed by the lossy qubit under microwave drive. The temperature of
the MR reaches a balance between the two baths as
\begin{equation}
T_{\text{eff}}=\frac{\Gamma T_0+\Gamma_1 T_1}{\Gamma+\Gamma_1}\ .
\end{equation}
The near-resonant microwave drive largely suppresses the fluctuation (see
Appendix)
\begin{equation}
\exp \left( \frac{\omega _{b}}{T_{1}}\right) \sim \left( \frac{1-\cos \beta
_{0}}{1+\cos \beta _{0}}\right) ^{2}\sim 1.
\end{equation}
Therefore we obtain $T_{1}\sim \omega _{b}\ll T_{0}\sim \Omega $,
which means the fluctuation caused by the qubit can be effectively
disregarded, as far as the spontaneous emission of the qubit is
concerned. A more realistic estimation of this fluctuation requires
detailed information about the quantum noise of the qubit.

The cooling of the MR could be measured by using the conventional
motion transduction method~\cite{Schwab2005pt}. In this system, it
can also be revealed by the reduction in the integration of the
power spectrum for the MR $\left\langle z^{2}\left( \omega \right)
\right\rangle $. Since
\begin{equation}
\delta I=\frac{2\sqrt{2}A\delta \omega }{A^{2}+2\delta \omega
^{2}}\frac{B_{0}I_{p}L_{0}}{\Delta }\delta z,
\end{equation}
the motion power spectrum is proportional to the current power
spectrum. The MR power spectrum can be recorded by detecting the
current in the loop. The current in the flux qubit loop can be
recorded with a dc SQUID switching measurement or with a more
sophisticated phase sensitive dispersive readout such as Josephson
bifurcation measurement~\cite{Siddiqi2004}.

\section*{Acknowledgement}

The authors are very grateful to M. H. Devoret J. E. Mooij, H.
Nakano and A. Kemp for their helpful suggestions regarding the
experimental realizations of this proposal. YDW also thank Yong Li,
Fei Xue, P. Zhang, F. Marquardt and C. Bruder for fruitful
discussions with them. This work is partially supported by the JSPS
KAKENHI (No. 18201018 and 16206003).

\appendix

\section*{Appendix: The dissipative dynamics of the qubit}

The dissipative dynamics of the qubit is governed by the following
master equation
\begin{equation}
\dot{\rho}=-\frac{i}{\hbar }\left[H_q,\rho \right] +\mathcal{L}\rho
\label{m1}
\end{equation}
where $\rho $ is the reduced density matrix of the qubit, $H_q$ is
the qubit Hamiltonian in eq.~(\ref{H1}) and $\mathcal{L}$ is the
Liouvillian characterizing the influence of the environment.

As regards the qubit, we only consider the spontaneous emission with
rate $\Gamma _{0}$ and neglect the excitation and dephasing terms
because the qubit energy spacing is much larger than the environment
temperature and the qubit is biased close to the degeneracy point
\begin{equation}
\mathcal{L}\rho =\frac{\Gamma _{0}}{2}\left( 2\sigma _{-}\rho \sigma
_{+}-\rho \sigma _{+}\sigma _{-}-\sigma _{+}\sigma _{-}\rho \right).
\label{spont}
\end{equation}
By successively applying three unitary transformations to
eq.~(\ref{m1})~\cite{Yong} or making use of the Fermi golden
rule~\cite{Hauss2007}, we can readily obtain the master equation in
the interaction picture of the rotating frame
\begin{equation}
\dot{\rho}_{I}^{R}=\mathcal{L}^{R}\rho _{I}^{R},
\end{equation}
where
\begin{eqnarray}
\mathcal{L}^{R}\rho _{I}^{R} &=&\frac{\Gamma _{\downarrow
}}{2}\left( 2\tilde{ \sigma}_{-}\rho _{I}^{R}\tilde{\sigma}_{+}-\rho
_{I}^{R}\tilde{\sigma}_{+}
\tilde{\sigma}_{-}-\tilde{\sigma}_{+}\tilde{\sigma}_{-}\rho
_{I}^{R}\right)  \notag \\
&&+\frac{\Gamma _{\uparrow }}{2}\left( 2\tilde{\sigma}_{+}\rho _{I}^{R}
\tilde{\sigma}_{-}-\rho _{I}^{R}\tilde{\sigma}_{-}\tilde{\sigma}_{+}-\tilde{
\sigma}_{-}\tilde{\sigma}_{+}\rho _{I}^{R}\right)  \notag \\
&&+\frac{\Gamma _{\varphi }}{2}\left( \tilde{\sigma}_{z}\rho _{I}^{R}\tilde{
\sigma}_{z}-\rho _{I}^{R}\right),
\end{eqnarray}
with
\begin{eqnarray}
\Gamma _{\downarrow } &=&\frac{\Gamma _{0}}{4}\left( 1+\cos \beta \right)
^{2},  \notag \\
\Gamma _{\uparrow } &=&\frac{\Gamma _{0}}{4}\left( 1-\cos \beta \right) ^{2},
\notag \\
\Gamma _{\varphi } &=&\frac{\Gamma _{0}}{2}\sin ^{2}\beta
\end{eqnarray}
and $\Gamma _{0}$ is the spontaneous emission rate. This master equation
shows that the effective bath formed by the driven qubit in dissipative
environment is
\begin{equation}
\exp \left( \frac{\omega _{b}}{T_{1}}\right)\equiv \frac{ \Gamma
_{\downarrow }}{\Gamma _{\uparrow }}=\left( \frac{1+\cos \beta _{0}}{1-\cos
\beta _{0}}\right) ^{2}
\end{equation}

Therefore, in the interaction picture the equations of motion for the
average value of the qubit operators are
\begin{eqnarray}
\frac{d}{dt}\left\langle \tilde{\sigma}_{+}\left( t\right)
\right\rangle &=&-\frac{\left( \Gamma _{\uparrow }+\Gamma
_{\downarrow }+2\Gamma _{\varphi }\right) }{2}\left\langle
\tilde{\sigma}_{+}\left( t\right) \right\rangle,
\notag \\
\frac{d}{dt}\left\langle \tilde{\sigma}_{-}\left( t\right)
\right\rangle &=&-\frac{\left( \Gamma _{\uparrow }+\Gamma
_{\downarrow }+2\Gamma _{\varphi }\right) }{2}\left\langle
\tilde{\sigma}_{-}\left( t\right) \right\rangle,
\notag \\
\frac{d}{dt}\left\langle \tilde{\sigma}_{z}\left( t\right) \right\rangle
&=&-\left( \Gamma _{\uparrow }+\Gamma _{\downarrow }\right) \left\langle
\tilde{\sigma}_{z}\left( t\right) \right\rangle -\left( \Gamma _{\uparrow
}-\Gamma _{\downarrow }\right).  \label{ds}
\end{eqnarray}
Setting $d\langle\sigma_i\rangle/dt=0$, we can obtain the
steady-state solution of the qubit in a dissipative system
\begin{eqnarray}
\left\langle \tilde{\sigma}_{+}\right\rangle _{s} &=&\left\langle
\tilde{\sigma}_{-}\right\rangle _{s}=0 ,  \notag \\
\left\langle \tilde{\sigma}_{z}\right\rangle _{s} &=&-\frac{\Gamma
_{\uparrow }-\Gamma _{\downarrow }}{\Gamma _{\uparrow }+\Gamma
_{\downarrow }}=-\frac{2\cos \beta \left( z\right) }{1+\cos
^{2}\beta \left( z\right) }. \label{s1}
\end{eqnarray}
Making use of eq. (\ref{s1}) and the transformation relations eqs.
(\ref{t1}), (\ref{t2}) and (\ref{t3}), it turns out that the
quantity of interest in the experimental frame is related to those
in the rotating frame as
\begin{equation}
\left\langle \sigma _{z}\right\rangle _{s}=\left( \cos \theta \left(
z\right) \cos \beta \left( z\right) -\sin \theta \left( z\right)
\sin \beta \left( z\right) \cos \omega _{d}t\right) \left\langle
\tilde{\sigma}_{z}\right\rangle _{s}.
\end{equation}
Neglecting the high-frequency oscillating term, we obtain the
effective value in eq. (\ref{sz})
\begin{equation}
\left\langle \sigma _{z}\right\rangle _{s}=-\frac{2\cos^{2}\beta
\left( z\right) }{1+\cos ^{2}\beta \left( z\right) }\cos \theta
\left( z\right)
\end{equation}
\qquad


\end{document}